\documentstyle[aaspp4]{article}
\begin{document}

\title{GLOBULAR CLUSTER FORMATION TRIGGERED BY THE INITIAL STARBURST 
       IN GALAXY FORMATION}

\author{Yoshiaki Taniguchi$^1$, Neil Trentham$^2$, \& Satoru Ikeuchi$^3$}

\affil{$^1$Astronomical Institute, Graduate School of Science, Tohoku University, 
       Aramaki, Aoba, Sendai 980-8578, Japan}
\affil{$^2$Institute of Astronomy, University of Cambridge, Madingley Road, Cambridge
       CB3 0EZ, UK}
\affil{$^3$Physics Department, Graduate School of Science, Nagoya University,
       Chikusa, Nagoya 464-8602, Japan}

\begin{abstract}
We propose and investigate
a new formation mechanism for globular clusters
in which they form within molecular clouds that are formed
in the shocked regions created
by galactic winds driven by successive supernova
explosions shortly after the initial burst of massive star 
formation in the galactic centers.
The globular clusters have a radial distribution
that is more extended than that of the stars
because the clusters form as
pressure-confined condensations in a shell that
is moving outward radially at high velocity.
In addition the model is consistent with existing
observations of other global properties of globular clusters, as
far as comparisons can be made.
\end{abstract} 


\section{INTRODUCTION}

Globular clusters (GCs) provide important clues as to how galaxies formed.
The very old ages of the GCs in our Galaxy suggest 
that they were formed when the Galaxy formed.  This, combined with the
fact that GCs  are normally found in the halos of big galaxies,
where the dark matter dominates the gravitational potential, implies that
the formation mechanism of GCs is related intimately to the
formation of the galaxies themselves. Therefore understanding the formation of
GCs has been a major area of study (e.g., van den Bergh 1996;
VandenBerg, Bolte, \& Stetson 1996; Harris et al. 1998). 

Globular clusters probably form within 
giant molecular clouds, in the same way as we see star cluster formation 
happening today in the Galactic disk (Harris \& Pudritz 1994;
McLaughlin \& Pudritz 1996; Elmegreen \& Efremov 1997).  
Indeed, there is considerable observational
support for such an approach, and the specific model of
McLaughlin \& Pudritz (1996) has the strong 
attraction of correctly reproducing the GC 
mass distribution function.
Therefore, the most important question now becomes:
{\it How were such giant molecular clouds formed earlier on in the 
history of the galaxies, well away from the galaxy centers which is where
most of the stars are being made?}
Proposed mechanisms for making molecular clouds include:
1) gravitational instability shortly after recombination arising from
isothermal perturbations in the early Universe (Peebles \& Dicke 1968;
Rosenblatt, Faber, \& Blumenthal 1988); 2) instabilities during
contraction of a protogalactic gas cloud (e.g., Fall \& Rees 1985, who
investigate thermal instabilities); and
3) high-velocity collisions of giant gas clouds in the halos of young galaxies
(Gunn 1980; Kang et al. 1990; Kumai, Basu, \& Fujimoto 1993).

Recent observations show evidence for candidate forming GCs, which
presumably formed out of molecular clouds,
in some galaxy mergers, suggesting that the third possibility is the most
attractive (Ashman \& Zepf 1992; Holtzman et al. 1992;
Zepf \& Ashman 1993;  Kumai et al.~1993, Whitmore et al.~1993,
Surace et al.~1998).
Such a scenario is further supported by the bimodal metallicity distributions
of the globular cluster populations in ellipticals (Zepf \& Ashman 1993)
and by the fact that mergers are
thought to produce elliptical galaxies by violent relaxation (Schweizer 1982)
in conjunction with the result that
elliptical galaxies have a higher specific frequency ($S_N$)
of GCs than disk galaxies (Zepf \& Ashman 1993).
However, the newly formed clusters are usually observed in the central
regions of mergers (the exception is in VII Zw 031, where star clusters have
been seen at ultraviolet wavelengths in a coherent pattern
at $\sim 5$ kpc from the galactic center;
Trentham, Kormendy, \& Sanders 1999), 
so that the mechanism producing these particular objects is
presumably not responsible for making most old globular cluster populations
located in the outer regions of galaxies.
While these observations suggest that halo GCs did not form in the current
merger, they do not rule out the possibility that they formed in high-velocity
cloud-cloud collisions early in the histories of the progenitor galaxies,
and therefore do not address 
directly the question highlighted in the previous
paragraph.

In the context of the models that we characterize 2) and 3) above, it is
plausible that any intense star formation that is happening in the galaxy
centers can have significant effects on the physical processes responsible
for the formation of the molecular clouds.
Recently, Harris et al. (1998) argued the importance of 
a superwind driven by an initial starburst in order to explain the observed
higher $S_N$ of GCs in bright cluster member galaxies;
the superwind is necessary to reduce the mass of cold gas
in which star formation occurs, leading to the higher $S_N$ (Blakeslee 1997).
In such a scenario the GCs would form as condensations in material shocked
by supernovae.  This concept is not new
(Mestel 1965; Elmegreen \& Lada 1977; Elmegreen \&
Elmegreen 1978; Elmegreen 1989; Whitworth et al.~1994;
Taniguchi, Trentham \& Shioya 1998; Mori, Yoshii \& Nomoto 1999), and we
now consider its application to the GC problem (see Fig.~1).  
That the GCs end up in the outer parts of the galaxy in this kind of model happens
because they form as pressure-confined condensations within a
shell of shocked material that is moving radially outward at high velocity.
This model exploits some features of 2) and 3), but differs fundamentally
from those listed above in that the formation of the halo GCs is related to the
formation of the dense stellar core of the galaxy, and that the core forms
early in the history of the galaxy, at least in elliptical galaxies. 
A similar approach has been undertaken by Brown, Burkert, \& Truran (1991,
1995), in which they consider GC formation in supershells produced by the
collective behavior of the supernovae remnants generated by the initial
starburst.  These models have had some success at reproducing the
properties of Milky Way GCs.

In the context of disk galaxies like the Milky Way, this model ties
halo globular cluster formation to bulge formation (note that bulges lie
on the same fundamental plane as the ellipticals $-$ 
see Kormendy \& Djorgovski 1989).  The physics is essentially
the same if the spheroidal stellar population (whether an elliptical galaxy or
the bulge of a disk galaxy) forms by isolated dissipative collapse
(Blumenthal et al. 1984) or by a merger-induced collapse (e.g. Kormendy
\& Sanders 1992).

\section{GLOBULAR CLUSTER FORMATION IN THE SHOCKED SHELL DRIVEN BY THE SUPERWINDS}

\subsection{Model}

First, we investigate if GCs could form in the shocked shell driven 
by a superwind caused by the initial starburst in a galaxy.
We adopt the dissipative collapse picture for formation of elliptical galaxies
and bulges (e.g.~Larson 1974) and follow the galactic wind model proposed by 
Arimoto \& Yoshii (1987). In this model,
intense star formation (i.e., a starburst) occurs at the epoch of
galaxy formation in the galaxy center,
producing a galactic wind which lasts for a characteristic
time $t_{\rm GW} (\sim 0.5$ Gyr for an elliptical with a stellar mass of
of $10^{11} M_\odot$).
Since infalling gas is accreting onto the galaxy
at times $t \geq t_{\rm GW}$, the wind interacts 
with this gas, 
and shocked gaseous shells form  
in the outer regions of the galaxy. If the shells are unstable
gravitationally (e.g.~Ostriker \& Cowie 1981; Ikeuchi 1981; 
Umemura \& Ikeuchi 1987),
clumps may be formed within them.
Here we investigate the possibility that the clumps may end up
as present-epoch GCs.  

Suppose that the supernovae responsible for shocking the gas occur continuously
over a timescale longer than or comparable with the dynamical timescale
of the initial gas cloud and  
the evolution of the shocked material can be described by a 
superbubble model
(McCray \& Snow 1979; Koo \& McKee 1992a, 1992b; 
Heckman et al. 1996; Shull 1995 and references therein).
The radius and velocity of the shocked shells at time $t$
(in units of 0.5 Gyr) are then

\begin{equation}
r_{\rm shell}  \sim 29 
L_{\rm mech, 43}^{1/5}
n_{\rm H, 1}^{-1/5}
t_{0.5}^{3/5} ~~ {\rm kpc},
\end{equation}
and

\begin{equation}
v_{\rm shell} \sim 34 
L_{\rm mech, 43}^{1/5}
n_{\rm H, 1}^{-1/5}
t_{0.5}^{-2/5} {\rm km ~ s}^{-1},
\end{equation}
where $L_{\rm mech}$ is the mechanical luminosity 
released collectively from the supernovae in the central starburst
in units of $10^{43}$ ergs s$^{-1}$ and $n_{\rm H}$ is the average 
hydrogen number density of the
ISM, assumed constant in units of 1 cm$^{-3}$.
The derivation of $r_{\rm shell}$ requires that the baryonic component
dominates the gravitational potential.  This is always true for
the relevant scales in this paper.  However the presence of a dark matter
halo requires that this estimate of $r_{\rm shell}$
is not valid at arbitrarily large radii.
We can estimate $L_{\rm mech}$ directly from Arimoto \& Yoshii (1987).
For an elliptical galaxy with a stellar mass 
$M_{\rm stars} = 10^{11} M_\odot$, 
radius  $r \simeq $ 10 kpc and 
${n}_{\rm H} \sim 1$  
cm$^{-3}$ (see Saito 1979; Arimoto \& Yoshii 1987),
we expect 
$N_{\rm SN} \sim 3 \times 10^9$ stars that explode as supernovae.
Since most of these massive stars were formed during the first 0.5 Gyr
(= $t_{\rm GW}$), therefore 
$L_{\rm mech} \sim \eta ~ E_{\rm SN} ~ N_{\rm SN} / t_{\rm GW} \sim 10^{43} ~ 
{\rm erg~ s}^{-1}$ 
where $E_{\rm SN}$ is the 
total energy of a single supernova
($10^{51}$ erg) and 
$\eta$ is the efficiency of the kinetic energy deposited to the ambient gas 
($\sim$ 0.1; Dyson \& Williams 1980).

Condensations that form within the shells experience a net inward
acceleration due to self-gravity and a net outward acceleration due to
the internal pressure.  Whitworth et al.~(1994) investigate the
balance between these two accelerations and show that the timescale
for the growth of the fastest-growing condensations is 
$t_{\rm fastest} \sim 2 c_{\rm s}/(G \Sigma)$, 
where $c_s$ is the sound speed in the shell, and
$\Sigma$ is its surface density.
Non-linear fragmentation in the shell then happens first at a time
$t = t_{\rm fastest}$.  Noting that the surface-density 
$\Sigma = C {{n}_{\rm H} } m_{\rm H} r_{\rm shell}$, 
where $C$ is a constant determined by the geometry ($C = 1/3$ for
a sphere) and $m_{\rm H}$ is the hydrogen atom mass,
we then find from the estimates of $r_{\rm shell}$ and
$v_{\rm shell}$ that fragmentation
within the shell first happens at a time 
$t_{\rm c} \simeq 57   
C_{0.33}^{-1/2}
n_{\rm H, 1}^{-1/2}
{\cal M}_{\rm c, 10}^{-1/2}$ Myr
at a radius
$r_{\rm c} \simeq  8 
L_{\rm mech, 43}^{1/5}
C_{0.33}^{-3/10}
n_{\rm H, 1}^{-1/2}
{\cal M}_{\rm c, 10}^{-3/10}$ kpc.
Here ${\cal M}_{\rm c, 10}$ is the Mach number in units of 10 when these
condensations first appear, equal to $v_{\rm shell}/c_{\rm s}$
(here it is assumed that ${\cal M}_{\rm c} \gg 1$; Whitworth et al.~1994).
Estimating $c_{\rm s}$ is difficult because the turbulent pressure in
the shell is much greater than the thermal pressure. 
The shell is moving outward at a velocity $v_{\rm c}$ when
the fragments first appear, where 
$v_{\rm c} \simeq 81
L_{\rm mech, 43}^{1/5}
C_{0.33}^{1/5}
{\cal M}_{\rm c, 10}^{1/5}$ km s$^{-1}$.
Thus $v_{\rm c}$ is almost independent of $c_{\rm s}$, so that our lack of
knowledge of the sound speed is unimportant in determining the
velocity of the shell when the condensations form.
 
Following Whitworth et al.~(1994), we can estimate the mass and size of
the fragments:
$M_{\rm frag} \sim {c_{\rm s}^{7/2}}/(G^3 n_{\rm H} m_{\rm H} v_{\rm c})^{1/2} 
\sim  3.8 \times 10^6 n_{\rm H, 1}^{-1/2}
v_{\rm c, 81}^{3}
{\cal M}_{\rm c, 10}^{-7/2} ~ M_\odot$ and
$l_{\rm frag} \sim c_{\rm s}^{3/2}/(G n_{\rm H} m_{\rm H} v_{\rm c})^{1/2}
\sim  247 n_{\rm H, 1}^{-1/2}
v_{\rm c, 81}
{\cal M}_{\rm c, 10}^{-3/2}$ pc
where $v_{\rm c, 81}$ is in units of 81 km s$^{-1}$.

Now we consider the evolution of the fragments within the shocked shell.
Since the cooling timescale of neutral gas clouds,
$t_{\rm cool} \sim 1 n_{\rm H,1}^{-1}$ Myr
(Spitzer 1978), once fragmentation has occurred,
the fragments will cool and can form stars.
It is not clear, however, whether such stars really evolve to form a
star cluster, or become smoothly distributed throughout
the galaxy
(e.g., Fall \& Rees 1985).
But one of the most important results of Whitworth et al.~(1994) was 
that fragmentation of the shocked layer occurs while 
the fragments are still confined within the layer by ram pressure.
Subsequent fragmentation within a fragment could occur and 
result in the formation of sub-GC clouds with Jeans (1929) masses of 
$m_{\rm J} \sim \lambda_{\rm J}^3 \rho_{\rm frag} \sim 3 \times
10^4 M_\odot$ for 
$\lambda_{\rm J} = c_{\rm s}^{\rm local} (\pi / G \rho_{\rm frag})^{1/2}
\sim 37 ~ (c_{\rm s}^{\rm local}/{\rm 1 ~ km ~ s^{-1}})$ pc and 
$\rho_{\rm frag} = M_{\rm frag}/[(4 \pi/3) (l_{\rm frag}/2)^3]
\sim 4 \times 10^{-23}$ g cm$^{-3}$.
The number of such sub-GC clouds is $N_{\rm sub} 
= M_{\rm frag}/m_{\rm J} \simeq 14$.
The lack of strong clustering amongst GCs in the halo suggests that these
sub-GC clouds merge within a condensation to form a single GC. 
Such merging happens on 
a dynamical timescale of 
$T_{\rm dyn} \sim N_{\rm sub}
l_{\rm frag}^{3/2} G^{-1/2} M_{\rm frag}^{-1/2}
\sim 8.3 \times 10^8 l_{\rm frag, 250}^{3/2} M_{\rm frag, 6}^{-1/2}$ yr,
where $l_{\rm frag, 250}$ is the original size of the fragment in units of
250 pc and $M_{\rm frag, 6}$ is the mass of the fragment in units
of $10^6 M_\odot$.
Since typical masses of GCs in the present-day galaxies are
$M_{\rm GC} \sim 10^5 M_\odot$, and these GCs are gas-poor,
about 90\% of the gas 
must have been removed from the initial fragments before virialization (such
a large fraction of gas being lost would unbind the system, if it happens
after virialization).
Supernova-driven winds could be an important mechanism in achieving this over
the lifetime of the GC; note that stars with masses above
$ 0.8 M_\odot$ in GCs have all evolved of the main-sequence by the present
day. 

Finally we estimate the location of GCs. The shocked shell
is confined on opposite sides by the ram pressure of the
inflowing ambient gas and by the hydrostatic pressure of the
expanding bubble. This results in the fragments being carried out
in the shocked layer well beyond $r_{\rm c}$. 
Pressure confinement ceases when the external pressure becomes
less than $G \Sigma^2$ i.e.~when
$t \simeq 0.18 C_{0.33}^{-1}
n_{\rm H, 1}^{-1/2}$ Gyr
and the maximum radial distance, $r_{\rm max} \simeq 16$ kpc.
Therefore the condensation leaves the shell at some radius $r$ where

\begin{equation}
8 L_{\rm mech, 43}^{1/5}
C_{0.33}^{-3/10}
n_{\rm H, 1}^{-1/2}
{\cal M}_{\rm c, 10}^{-3/10} ~
{\rm kpc}  <  r 
< 16 L_{\rm mech, 43}^{1/5}
C_{0.33}^{-3/5}
n_{\rm H, 1}^{-1/2} ~~ {\rm kpc}.
\end{equation}
This is much larger than the stellar half-light radii of elliptical
galaxies and bulges (Kormendy \& Djorgovski 1989). 
Subsequent dynamical evolution of
the GCs will depend on the gravitational potential at these large radii,
which progressively becomes more dark-matter dominated as the halo
virializes.

\subsection{Confrontation with Observation} 

The main result following from the previous section is that the remnant
stellar clusters have galactocentric radii between 8 and 16 kpc, for an
initial starburst of luminosity 10$^{43}$ erg s$^{-1}$.  These
numbers are approximately consistent with the galactocentric radii of
globular clusters in our Galaxy (5 -- 10 kpc; Harris 1991, 
Harris et al.~1998).  The radii in our Galaxy could be lower than those
inferred from the model for various reasons: for example, dynamical friction
might reduce the size of the orbits of the GCs, or the initial starburst
in our Galaxy might have generated a luminosity lower than
10$^{43}$ erg s$^{-1}$.  

The following confrontations with observation can also be made.  The model
is highly idealized and it is probably too simplistic to merit some of the
more detailed comparisons.  Nevertheless some useful constraints on the
model parameters and some important extensions to the model can be inferred.
   
\noindent 1) Since the GCs are formed in the shocked shell
which is the interface between the metal-enriched
galactic wind and the metal-poor accreting gas, the most obvious
scenario would be that metallicity of the
GCs would be metal-poor, but slightly more metal-rich
than the lowest metallicity stars in the galaxy centers (the first ones
to form in the starburst that generated the superwind).
Alternatively, if supernovae from the central starburst inject many
metals into the expanding shell (Brown et al.~1991, 1995), then the
metallicity of the GCs will be much higher.
This comparison can only be made in the case of the Galaxy.
The median metallicity for halo GCs is about $-1.5$ in
logarithmic solar units.
The metallicities of stars in the bulge range from
about $-2$ to $1$ (Geisler, \& Friel 1992; McWilliam, \& Rich 1994).
This would suggest that the enrichment of the shell is not a highly
efficient process.
Furthermore, the formation epoch of GCs is shortly after that
of galaxy formation so that the ages of the globular cluster stars
should be nearly the same as the ages of the oldest stars in the galaxies.
This also appears to be true for the Galaxy.
 
\noindent 2) If the infalling gas has no angular momentum,
the GCs form in this model with highly radial orbits. 
Were this the case, and were these orbits to 
survive until the present day, this would be
inconsistent with observation, at least for the Galaxy
(van den Bergh 1993; Cohen \& Ryzhov 1997).
However, collisions with other condensations early in the history of the
fragments will randomize the orbits. 
This, combined with out lack of knowledge about the distribution of
angular momenta of the infalling gas, means that a detailed comparison with
observation is not possible.  

\noindent
3) One further consequence of our model is that {\it if} in all galaxies
the efficiency of globular 
cluster formation in the shocked shells is the same,
then the number of globulars $N_{\rm GC}$ scales as the luminosity $L$
of the galaxy as $N_{\rm GC} \sim r_{\rm c}^{3} \rho_{\rm c} M_{\rm frag}^{-1} 
\sim r_{\rm c}^{3} v_{\rm c}^{1/2} \sim L^{1/2}$, where $\rho_{\rm c}$
is the mass density of the GC clouds. 
Most data suggests a scaling law steeper than this (Harris 1991).
This may be due to that other physical processes may well be at work;
e.g., mergers and accretion events (Djorgovski \& Santiago 1992;
Ashman \& Zepf 1992; van den Bergh 1993; Zepf, Ashman,
\& Geisler 1995), and destruction of GCs as a result of 
evaporation, disk shocking, or dynamical
friction (Fall \& Rees 1977; Okazaki \& Tosa 1995).

\noindent 4) Local density variations (i.e.~different values of
$\Sigma$ in different regions) within the shell might mean
that there may be a radial
(and possibly age) spread amongst the globulars that
form by the mechanism described in the previous section.
In regions of high-$\Sigma$, the condition of 
$t_{\rm fastest} \sim 2 c_{\rm s}/(G \Sigma)$ is satisfied sooner,
so that the condensations begin to form within the shell earlier.
These systems also leave the shell earlier and so we expect them to
exist at systematically smaller galactocentric radii than systems
that form in the low-$\Sigma$ regions of the shell.  
Observations seems to indicate (Harris 1991) that halo GCs tend to be
smaller with increasing distance from the galactic centers.  In the
context of the current model, this would suggest that the efficiency at
which gas is converted into stars is higher in the high-$\Sigma$ regions. 

\vspace {0.5cm}

We would like to thank N. Arimoto, T. Murayama, Y. Ohyama, T. Okazaki,
H. Saio,  Y. Shioya, and M. Tosa
for their useful discussion and comments.
This work was supported by
the Ministry of Education, Science, and Culture
(Nos. 07044054, 10044052, and 10304013).
NT thanks the PPARC for financial support.


\newpage

\begin{figure}
\epsfysize=18.5cm \epsfbox{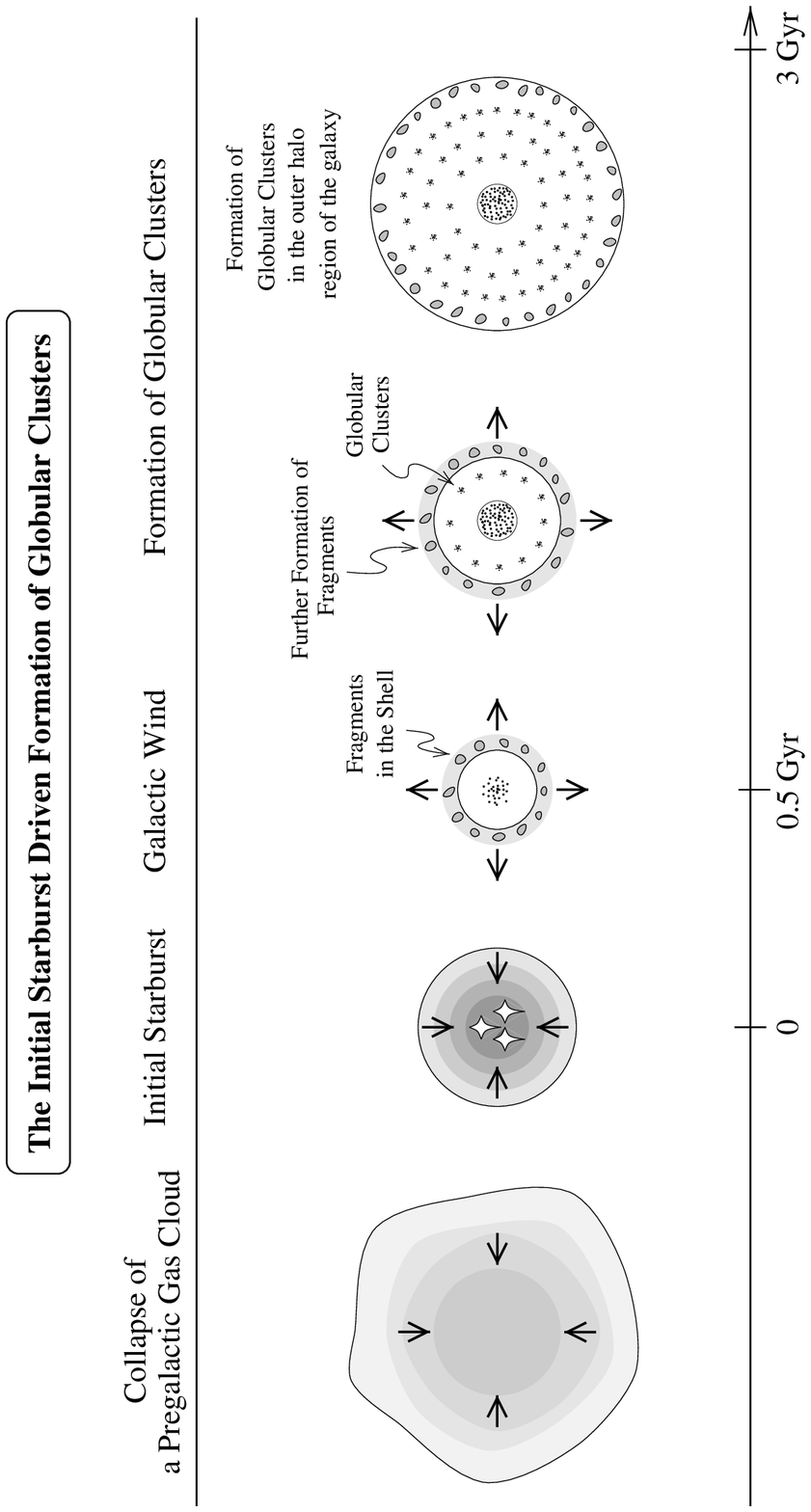}
\caption[]{
A schematic illustration of the initial starburst-driven
formation of globular clusters.
\label{fig1}
}
\end{figure}

\end{document}